\documentclass[12pt]{article}
\newcommand{\be}{\begin{equation}}
\newcommand{\ee}{  \end{equation}}
\newcommand{\ba}{\begin{eqnarray}}
\newcommand{\ea}{  \end{eqnarray}}

\usepackage{graphicx}
\usepackage{float}
\usepackage{xcolor}
\usepackage{bbm}
\usepackage{amsmath}
\usepackage{bbold}
\usepackage{mathtools}
\usepackage{dcolumn}
\usepackage{bm}
\usepackage{mathtools}
\usepackage{multirow}
\usepackage[colorlinks=true,linkcolor=blue,citecolor=blue,urlcolor=blue]{hyperref}
\usepackage{cleveref}

\usepackage{geometry}
\RequirePackage[normalem]{ulem}
\RequirePackage{color}

\title{Matrix H-theory approach to stock market fluctuations}
\author{Luan M. T. de Moraes and Ant\^onio M. S. Macedo¹,\\ Raydonal Ospina², \\Giovani L. Vasconcelos³}
\date{\vspace{-5ex}}
\newcommand{\affiliations}{\textit{¹Federal University of Pernambuco, Department of Physics.\\ ²Federal University of Bahia, Department of Statistics. \\³Federal University of Paran\'a, Department of Physics.}}
\begin{document}

\maketitle
\begin{center}
\affiliations
\end{center}

\begin{center}
\date{\today}
\end{center}
\begin{abstract}
We introduce matrix H theory, a framework for analyzing collective behavior arising from multivariate stochastic processes with hierarchical structure. The theory models the {joint distribution of the multiple variables (the measured signal)} as a compound of a large-scale multivariate {distribution}  {with the distribution of a  slowly fluctuating background. The  background is characterized by a hierarchical stochastic evolution of internal degrees of freedom, representing the correlations between stocks at different time
scales}. {As in its univariate version,} the matrix H-theory formalism also has two universality classes: Wishart and inverse Wishart, enabling a concise description of both the background and the signal probability distributions in terms of Meijer ${G}$-functions with matrix argument. Empirical analysis of daily returns of stocks within the S\&P500  demonstrates the effectiveness of matrix H theory {in describing fluctuations in stock markets}. These findings contribute to a deeper understanding of multivariate hierarchical processes and offer potential for developing more informed portfolio strategies in financial markets.

\end{abstract}

\section{Introduction}
Complex systems are characterized by a hierarchical organization of multiple spatial and temporal scales. The presence of distinct scales within the system's dynamics indicates the interplay of multiple processes operating at different levels of organization \cite{anderson1972more}. These scales can often be identified in multivariate time series data through the analysis of temporal correlations and cross-correlations across different variables \cite{PhysRevE.51.5084,referee22}.
The detailed analysis of such multivariate data has been significantly advanced in recent years by means of interdisciplinary approaches \cite{lutkepohl2005new,kachigan1986statistical} that find applications in various fields, including fluid mechanics \cite{beck2005time}, random lasers \cite{gomes2023levy}, neuroscience \cite{pereda2005nonlinear}, and econophysics \cite{mantegna1999introduction}. Notably, financial markets exhibit non-equilibrium properties where the collective behavior and correlations among assets—viewed as degrees of freedom—are crucial for both academic research and practical applications such as market prediction, hedging, leverage investments, and portfolio optimization.

Despite numerous approaches since the pioneering work of Bachelier \cite{bachelier1900theory}, the optimal model for describing financial markets remains a topic of debate. Several landmark studies have provided insights into market behavior, drawing on physical theories such as Brownian motion \cite{voit2003brownian}, cascade information flux \cite{ghashghaie1996turbulent}, and criticality/phase transitions \cite{bury2013statistical, bury2013market}. 
To account for the heavy tails often observed in financial time series \cite{mandelbrot1997variation}, stochastic volatility models have been proposed, addressing certain non-equilibrium effects albeit with a simple noise modeling for volatility \cite{hull2016options}. Works by Guhr and Kälber \cite{guhr2003new} and  Schmitt et al.~\cite{schmitt2013non}  have shown that random matrix theory (RMT) could effectively describe the covariance matrix of daily returns in financial markets. They posited that stock returns follow a Gaussian distribution modulated by a Wishart random matrix covariance noise, thereby explaining the heavy tails in long-term returns. Complementary studies  \cite{laloux1999noise, plerou1999universal} have shown that the spectrum of the covariance matrix of financial assets aligns well with the Marcenko-Pastur distribution, although some deviations were interpreted as signals of underlying information within the covariance matrix.
Noh \cite{noh2000model} introduced a model involving a block-diagonal covariance matrix to characterize correlations between asset sectors and random noise, addressing the discrepancies noted in earlier studies and attributing the observed signals to noise structured by RMT. 

More recently \cite{salazar2012multicanonical, macedo2017universality}, a  formalism known as H theory was introduced as a generic framework for understanding the emergence of non-Gaussian behavior from hierarchical stochastic processes. This formalism was found to describe well the occurrence of heavy-tailed fluctuations observed in various systems across diverse disciplines \cite{sosa2019emergence,barbosa2022turbulence, vasconcelos2024turbulence, gonzalez2017turbulence}. Notably, H-theory identifies two broad universality classes within these hierarchical processes, distinguished by their asymptotic behavior: power-law and stretched exponential tails.

In this paper, we aim to extend the hierarchical concepts of H-theory to multivariate time series by analyzing the stocks listed on the S\&P500 stock index. 
We assume that the covariance matrix is random but presents a hierarchical structure with different time scales and interaction among them, {while  the distribution of returns conditioned to a fixed covariance matrix is a multivariate Gaussian. The marginal distribution of returns is then obtained as a matrix compounding of this multivariate Gaussian with an appropriate  hierarchical family of distributions for the  covariance matrix}.  {Two  universality classes of ensembles for the covariance matrix are considered, namely  hierarchical Wishart and inverse Wishart, both of them yielding analytical formulas for the distributions of returns and covariances in terms of special functions (the Meijer $G$-functions) of matrix argument.}

{The matrix H-theory formalism introduced here allows one} to identify the number of relevant time scales and to determine the most suitable statistical model for multivariate time series.
 {As one of the main applied results reported in the present paper,} we demonstrate that the behavior of the return distribution for the S\&P500 stocks can be well described using the Wishart class, {in agreement with previous results on multivariate analysis of stock market fluctuations \cite{schmitt2013non,manolakis2023analysis}. However, contrary to these works (that intrinsically assume only one relevant time scale in the covariance matrix dynamics), we show that at least  three distinct time scales are necessary to faithfully describe the distributions of returns of the S\&P500 stocks.}

{The paper is structured as follows. Section II provides a summary of H theory in the univariate case and introduces its extension to matrix H theory. Section III discusses the unidimensional projections of multivariate distributions, drawing parallels with their univariate counterparts, and presents the methodology for real data application. Section IV revisits key findings from prior works, providing context and continuity. Section V applies the developed concepts to S\&P500 assets data, showcasing practical use cases. Section VI discusses the implications of matrix H theory and the insights gained from the analysis.}

\section{H Theory}

In this section we introduce the matrix version of an hierarchical model \cite{macedo2017universality}, but first we summarize the main aspects of the univariate case.

\subsection{Univariate H Theory: A Brief Review}

Here, for completeness, we present a brief summary of H theory for univariate time series, which will then be extended to the multivariate case in the following section.

We consider a time series $x(t)$ obtained from measurements of a multiscale  dynamical system. For example, in fluid turbulence the basic variables $x(t)$  of interest are the velocity increments computed at the (shortest) time scale of the data acquisition  \cite{castaing1990}, while in asset price dynamics one usually considers $x(t)$ to be the price logarithmic returns  at the (highest) frequency of the data \cite{vasconcelos2024turbulence}.  To be specific, we assume that there is a number,  $N$, of well-separated times scale, $\tau_i$,  between the shortest time scale, $\tau_N$, where the measurements  are made, and the largest time scale, $\tau_0$, above which  no correlations in the series would be present, so that  $\tau_{0}\gg \tau_{1}\gg \cdots \gg \tau_{N}$. The large-scale distribution,   $P_0(x/\sqrt{\varepsilon_0})$,  is  assumed to be known, where the  parameter $\varepsilon_0$ characterizes the `global equilibrium' of the system. The central hypothesis  of H theory \cite{macedo2017universality} is that the short scale distribution $P_N(x)$  is obtained from a compound of distributions:
\be\label{signal}
P_N(x)=\int P_0\left(\frac{x}{\sqrt{\varepsilon_N}}\right)f_N(\varepsilon_N)d\varepsilon_N,
\ee 
where  $f_N(\varepsilon_N)$ is the probability distribution of some `background' variable $\varepsilon_N$. 
As already mentioned, in the context of turbulence the variables $x(t)$ are velocity increments at some short time  scale, in which case  $\varepsilon_N$ can represent the energy flux to that scale \cite{macedo2017universality}. Similarly, in financial data the relevant signal $x(t)$ is usually  the  short-time logarithmic returns, so that $\varepsilon_N$ can be seen as the short-horizon volatility \cite{vasconcelos2024turbulence}.

Relation \eqref{signal} encodes the important information that over short periods of times the system tends to relax to a quasi-equilibrium whose distribution is of the same form as the large-scale distribution. In other words, the short-scale distribution conditioned  to a fixed background reproduces the large-scale distribution \cite{castaing1990,castaing1994}: $P_N(x|\varepsilon_N)=P_0(x/\sqrt{\varepsilon_N})$. However, owing to the coupling across nearby scales, the  background parameter $\varepsilon_N$ fluctuates in time, so one needs to integrate over all possible values of $\varepsilon_N$ to obtain the  marginal distribution  $P_N(x)$. 
Since at large scales the series $x(t)$ is usually uncorrelated, a natural choice for  $P_0(x/\sqrt{\varepsilon_0})$ is the Gaussian distribution (for similar arguments in financial markets, see \cite{_2001}):
\be 
P_0(x/\sqrt{\varepsilon_0})=\frac{1}{\sqrt{2 \pi \varepsilon_0}}\exp \left( -\frac{x^2
}{2 \varepsilon_0}\right).
\label{Gauss}
\ee 
In the H theory formalism the background distribution $f_N(\varepsilon_N)$ is obtained from a hierarchical model (rather than postulated as in other compounding-based approaches \cite{castaing1990,beck2005time}),  as discussed next.

The fluctuating background is modelled  by a set of coupled stochastic differential equations \cite{macedo2017universality}:
\be 
{d\varepsilon _i}(t)=-\gamma _i\left( \varepsilon _{i}-\varepsilon _{i-1}\right) dt+\kappa _i
\varepsilon_{i}^s\varepsilon_{i-1}^{1-s}dW_i(t). 
\label{eq:genSDE}
\ee 
for $i=1,...,N$, where $\gamma _i$ and $\kappa _i$ are positive constants and $s=1/2, 1$ (see below). {As discussed elsewhere \cite{macedo2017universality}, the form of Eq.~(\ref{eq:genSDE}) is dictated by three basic requirements, namely: i) a `global equilibrium' condition,  $\langle \varepsilon_i\rangle=\varepsilon_0$, $\forall i$, in the stationary regime ($t\to \infty$), where brackets indicate average; ii) positivity of $\varepsilon_i$, which requires that the noise amplitude vanishes for  $\varepsilon_i=0$; and iii) invariance under change of scale (i.e., $\varepsilon_i \to \lambda  \varepsilon_i$, for an arbitrary factor $\lambda$), so that the right hand side of (\ref{eq:genSDE}) must be a homogeneous function of degree one. Although for any exponent $0<s\le1$ these  requirements are fulfilled, we will see below that two values of $s$, namely $s=1/2$ and $s=1$, are special in that they lead to analytic solutions for the model in terms of certain higher transcendental functions (the Meijer $G$-functions).} 
{Note also that for fixed $\varepsilon_{i-1}$, Eq.~(\ref{eq:genSDE}) is a mean reverting process, hence $\langle \varepsilon_i|\varepsilon_{i-1}\rangle=\varepsilon_{i-1}$, indicating  the hierarchical nature of the process, whereby at any level $i$ of the hierarchy the variable $\varepsilon_i$ tends to `equilibrate' with the level immediately above.}

Holding the slower variable $\varepsilon_{i-1}$ fixed, one can compute the stationary  solution $f(\varepsilon_i|\varepsilon_{i-1})$ of the corresponding Fokker-Planck equation,  yielding two cases of interest. For $s=1/2$ one obtains the gamma distribution,
\be
 f(\varepsilon_i|\varepsilon_{i-1}) = \frac{{(\beta_{i}/ \varepsilon_{i-1})}^{\beta_{i}}}{\Gamma (\beta_{i}) } {\varepsilon_i^{\beta_{i}-1}}  e^{{-\beta_{i} \varepsilon_{i}}/{\varepsilon_{i-1}}}, \label{eq:gamma}
\ee
whilst for $s=1$ the inverse-gamma distribution follows:
\be
 f(\varepsilon_i|\varepsilon_{i-1}) = \frac{{(\beta_{i} \varepsilon_{i-1})}^{\beta_{i}+1}}{\Gamma (\beta_{i}+1) } {\varepsilon_i^{-\beta_{i}-2}}  e^{{-\beta_{i} \varepsilon_{i-1}}/{\varepsilon_i}}, \label{eq:invgamma}
\ee
where $\beta_i=2\gamma_i/\kappa_i^2$.

The  background distribution, $f_N(\varepsilon_N)$,  is then  obtained by integrating over the intermediate scales:
\be\label{background}
f_N(\varepsilon_N)=\int d\varepsilon_1 \cdots \int  d\varepsilon_{N-1} \prod_{i=1}^{N}f(\varepsilon_i|\varepsilon_{i-1}).
\ee
The integrals in (\ref{background}) can be performed exactly using the convolution theorem of the Mellin transform \cite{debnath2016integral} and the result can be written in terms of the Meijer $G$-functions $G^{ m,n}_{ p,q}$ (see Appendix \ref{meijer G appendix}).
For the gamma class ($s=1/2$) we find
\begin{equation}
\label{gamma}
    f_N(\varepsilon_N )=
\frac{\omega}{\varepsilon_0\Gamma(\boldsymbol \beta)}    G_{ 0,N } ^{ N,0 }  \left( 
\begin{array}{c}
{-} \\ 
{ \boldsymbol\beta-{\bf 1}}
\end{array}
\bigg |\frac{\omega \varepsilon_N}{\varepsilon_0 }  \right),
\end{equation}
where $\omega  =\prod_{j=1}^{N}\beta _j$ and we have introduced the vector notations
${\boldsymbol\beta}\equiv (\beta_1,\dots,\beta_N)$ and 
$
\Gamma({\bf a})  \equiv\prod_{j=1}^{N}\Gamma (a _j).
$
Similarly, for the inverse-gamma class ($s=1/2$) we get
\begin{equation}
\label{meijer1}
    f_N(\varepsilon_N )= \frac{1}{\varepsilon_0\omega\Gamma(\boldsymbol\beta+{\bf 1})}    G_{ N,0 } ^{ 0,N }  \left( 
\begin{array}{c}
{- \boldsymbol\beta-{\bf 1}}\\ 
-
\end{array}
\bigg |\frac{ \varepsilon_N}{\varepsilon_0 \omega} \right).
\end{equation}

Now, inserting (\ref{Gauss}) and (\ref{gamma}) into (\ref{signal}), and performing the integral  we get for the gamma class
\be 
P_N(x)=
\frac{\omega^{1/2}}{\sqrt{2\pi\varepsilon_0}\Gamma(\boldsymbol\beta)}G_{0,N+1}^{N+1,0}\left( 
\begin{array}{c}
- \\ 
{ \boldsymbol\beta-{\bf 1/2}},0
\end{array}
\bigg |\frac{\omega x^2}{2\varepsilon_0}\right).
\label{eq:PN2}
\ee 
The tail of this distribution  is given by a modified stretched exponential
\be 
P_N(x)\sim 
{x^{2\theta}}{\exp\left[-(N+1)(\omega x^2/2\varepsilon_0)^{1/(N+1)}\right]},\; |x|\to\infty,
\label{eq:asym2}
\ee 
 where $\theta=(\sum_{i=1}^N\beta_i -N)/(N+1)$.
Similarly, for the inverse-gamma class we obtain
\be 
P_N(x)=
\frac{1}{\sqrt{2\pi\omega\varepsilon_0}\Gamma(\boldsymbol\beta+{\bf 1})}G_{N,1}^{1,N}\left( 
\begin{array}{c}
{ -\boldsymbol\beta-{\bf 1/2}} \\ 
0
\end{array}
\bigg |\frac{x^2}{2\omega \varepsilon_0}\right),
\label{eq:PN1}
\ee 
which has power-law tails:
\be 
P_N(x)\sim 
 \sum_{i=1}^{N} \frac{c_{i}}{x^{2\beta_i+3}}, \quad  \quad |x|\to\infty,
 \label{eq:asym1}
\ee 
where the $c_i$ are constants (see \cite{boucheron2015tail,macedo2017universality,dufresne2010g} for more details).
\par
One can show that in the limit of a large number of scales, the lognormal background distribution emerges in all classes, thus recovering an expected behavior, e.g.,  for the distribution of energy dissipation in intermittent turbulence (in the limit of very high Reynolds number), as originally proposed by Obukhov \cite{Obukhov1962} and Kolmogorov \cite{Kolmogorov1962}; see also \cite{castaing1994}. To be more specific, it can be shown that

\be
\lim_{N,\beta_i\to\infty} f_N(\varepsilon) = \frac{1}{\varepsilon\sigma\sqrt{2\pi}} \exp\left(-\frac{(\ln \varepsilon - \lambda)^2}{2\sigma^2}\right), 
\ee
where \(\lambda\) and \(\sigma\) are constants. Intuitively, this can be seen by introducing the variable
\(\varepsilon = \xi_N\xi_{N-1}\dots\xi_1\),
where \(\xi_i = \varepsilon_i/\varepsilon_{i-1} \) and realizing that the variables \(\xi_i\) are statistically independent when the time scales are largely separated. Finally, using the central limit theorem one can establish that \(\ln \varepsilon = \sum_{i=1}^N \ln \xi_i\) has a Gaussian distribution.

Next, we will see how the H theory formalism summarized above can be nicely extended to the multivariate case.
 
\subsection{Multivariate Case} 
\bigskip
\par
Let $\boldsymbol{r}^\top=(r_1,r_2,...,r_p)$ be a   random vector in $\mathbbm{R}^p,$ where the superscript $\top$ stands for transpose. For example, each random variable $r_i$ may represent the returns of a given stock computed at some short time scale $\tau_N$ (say, daily or intraday returns). The set of $p$ companies considered may correspond, for instance, to  companies from a given sector of the economy or companies that enter a given stock exchange index.

As in the univarite case, the joint distribution, $P_N(\boldsymbol{r})$, of short-scale returns  is  written as
\be\label{signal2}
P_N(\boldsymbol{r})=\int P(\boldsymbol{r}|\Sigma_N)f_N(\Sigma_N)d\Sigma_N,
\ee 
where the conditional distribution 
is assumed to be a multivariate Gaussian distribution,
\be\label{Gauss2}
P(\boldsymbol{r}|\Sigma_N)=\frac{1}{|2\pi\Sigma_N|^{1/2}}\exp\left( -\frac{1}{2}\boldsymbol{r}^\top\Sigma_N^{-1}\boldsymbol{r}\right),
\ee
and $f_N(\Sigma_N)$ is probability density of the short-scale covariance matrix $\Sigma_N$. Here $\Sigma_N$ is a $p\times p$ real, symmetric, positive definite matrix and $|\Sigma_N|\equiv\det(\Sigma_N)$. 
 
As will be  shown in Sec.~\ref{applicationSection}, the financial data used here supports a Gaussian description (for fixed background) at short time scales.

As before, the multivariate background density $f_N(\Sigma_N)$ is obtained from a hierarchical series of convolutions:
\be\label{background2}
f_N(\Sigma_N)=\int  d\Sigma_1 ...d\Sigma_{N-1} \prod_{i=1}^{N}f(\Sigma_i|\Sigma_{i-1}),
\ee
where $f(\Sigma_i|\Sigma_{i-1})$ is the background density at scale $\tau_i$ for fixed $\Sigma_{i-1}$ and $\Sigma_0$ is the fixed covariance matrix of the largest scale.  As in the univariate case, we must also choose $\langle \Sigma_i|\Sigma_{i-1}\rangle=\Sigma_{i-1}$, to capture the hierarchical nature of the background dynamics.
Analogously to the gamma and inverse-gamma classes of the univariate formalism, explicit expressions can be obtained for two classes: (i) Wishart and (ii) Inverse-Wishart. We shall consider them separately.

\subsubsection{Wishart Class}

This class is defined by the choice of $f(\Sigma_i|\Sigma_{i-1})$ as a Wishart distribution
\ba \label{wishartdistribution}
f(\Sigma_i|\Sigma_{i-1})&=&\frac{|\beta_i\Sigma_{i-1}^{-1}|^{\beta_i}}{\Gamma_p(\beta_i)}|\Sigma_i|^{\beta_i-(p+1)/2} \nonumber \\ &&\times \; \exp\left(-\beta_i{\rm Tr}\left( \Sigma_{i-1}^{-1}\Sigma_i\right)\right),
\ea
where {$\beta_i>0$ are free parameters and}
\be \label{multivariategamma}
\Gamma_p(\beta)\equiv\pi^{p(p-1)/4}\Gamma(\beta)\Gamma(\beta-\tfrac{1}{2})\Gamma(\beta-1)\cdots\Gamma(\beta-\tfrac{(p-1)}{2})
\ee
is the matrix-variate gamma function \cite{mathai1991multivariate}. Just as in the univariate case, the multiple integral (\ref{background2}) can be calculated using a matrix version of the Mellin transform, defined as \cite{mathai1997jacobians}
\be\label{MTransformMatrix}
M[f;s] \equiv \int_{X>0} dX |X|^{s-(p+1)/2}f(X),
\ee 
where $X$ is $p\times p$ real, symmetric, positive definite matrix. The integral in (\ref{background2}) can  be expressed in terms of the Meijer $\bar{G}$-function of matrix argument, denoted by
\be
\bar{G}(X)=\bar{G}_{p,q}^{m,n}\left( 
\begin{array}{l}
\boldsymbol{a} \\ 
\boldsymbol{b}
\end{array}
\bigg | X\right) \label{Matrix-Mellin-G},
\ee 
where $\boldsymbol{a}=(a_1,\ldots ,a_p)$ and $\boldsymbol{b}=(b_1,\ldots ,b_q)$,
and  defined through its (matrix) Mellin transform (see Appendix \ref{appendixproperties}):
\ba
M[\bar{G};s]&=&\frac{\prod_{j=1}^m\Gamma_p
(s+b_j)}{\prod_{j=m+1}^q\Gamma_p
((1+p)/2-s-b_j)} \nonumber\\
&&\times \; \frac{\prod_{j=1}^n\Gamma_p ((1+p)/2-s-a_j)}{\prod_{j=n+1}^p\Gamma_p (s+a_j)}.
\label{eq:MbarG}
\ea
(Here we use $\bar G$ to denote the matrix-argument Meijer $G$-function, so as to avoid confusion with the standard one-variable $G$-function.) 

Inserting (\ref{wishartdistribution}) into (\ref{background2}), and using  properties of the Mellin transform, we find  
\begin{equation}
\label{gamma2}
    f_N(\Sigma_N )=
\frac{|\omega \Sigma_0^{-1}|^{(p+1)/2}}{\Gamma_p(\boldsymbol \beta)}   \bar{G}_{ 0,N } ^{ N,0 }  \left( 
\begin{array}{c}
{-} \\ 
{ \boldsymbol\beta-\frac{p+1}{2}{\bf 1}}
\end{array}
\bigg |\frac{\omega \Sigma_N}{\Sigma_0 }  \right),
\end{equation}
Now, inserting (\ref{gamma2}) into (\ref{signal2}) we get
\ba\label{int1}
P_N(\boldsymbol{r})&=&\frac{|\omega \Sigma_0^{-1}|^{(p+1)/2}}{\Gamma_p(\boldsymbol \beta)}\int d\Sigma_N P(\boldsymbol{r}|\Sigma_N) \nonumber \\ &&\times \; \bar{G}_{ 0,N } ^{ N,0 }  \left( 
\begin{array}{c}
{-} \\ 
{ \boldsymbol\beta-\frac{p+1}{2}{\bf 1}}
\end{array}
\bigg |\frac{\omega \Sigma_N}{\Sigma_0 }  \right).
\ea
To perform the integral, we start by using a Hubbard-Stratonovitch-transformation \cite{sorella1991hubbard} to represent $P(\boldsymbol{r}|\Sigma_N)$ as \be\label{HS}
P(\boldsymbol{r}|\Sigma_N)=\int \frac{d\boldsymbol{k}}{(2\pi)^p}e^{-i\boldsymbol{k}\cdot\boldsymbol{r}}\exp \left( -\frac{1}{2}\boldsymbol{k}^\top\Sigma_N\boldsymbol{k}\right),
\ee
where $\boldsymbol{k}=(k_1,...,k_p)$. Inserting (\ref{HS}) into (\ref{int1}), we get
\be\label{int2}
P_N(\boldsymbol{r})=\int \frac{d\boldsymbol{k}}{(2\pi)^p}e^{-i\boldsymbol{k}\cdot\boldsymbol{r}}I(\boldsymbol{k}),
\ee
where
\ba
I(\boldsymbol{k})&=&\frac{|\omega \Sigma_0^{-1}|^{(p+1)/2}}{\Gamma_p(\boldsymbol \beta)}\int d\Sigma_N \exp \left( -\frac{1}{2}\boldsymbol{k}^\top\Sigma_N\boldsymbol{k}\right) \nonumber \\ &&\times \; \bar{G}_{ 0,N } ^{ N,0 }  \left( 
\begin{array}{c}
{-} \\ 
{ \boldsymbol\beta-\frac{p+1}{2}{\bf 1}}
\end{array}
\bigg |\frac{\omega \Sigma_N}{\Sigma_0 }  \right).
\ea
We define the new variable $X=(\omega \Sigma_0^{-1})^{1/2}\Sigma_N(\omega \Sigma_0^{-1})^{1/2}$, so that $dX=|\omega \Sigma_0^{-1}|^{(p+1)/2}d\Sigma_N$. Using the symmetry of the $\bar{G}$ function of matrix argument (in the sense $\bar{G}(AB) = \bar{G}(BA)$),  we may rewrite $I(\boldsymbol{k})$ as
\ba \label{CFTWishart1}
I(\boldsymbol{k})&=&\frac{1}{\Gamma_p(\boldsymbol \beta)}\int dX \exp \left( -\frac{1}{2\omega}\boldsymbol{k}^\top\Sigma_0^{1/2}X\Sigma_0^{1/2}\boldsymbol{k}\right) \nonumber \\ &&\times\;\bar{G}_{ 0,N } ^{ N,0 }  \left( 
\begin{array}{c}
{-} \\ 
{ \boldsymbol\beta-\frac{p+1}{2}{\bf 1}}
\end{array}
\bigg |X   \right)
\ea

The next step is to use the rotation invariance of random matrix ensembles and perform a ``color-flavor-type" transformation (CFT) [see Eq.~(\ref{Meijer-Wishart CFT}) in Appendix \ref{Meijer-Wishart CFT subsection}], {with which we map the multivariate integral onto its univariate ($p=1$) version}. Thus we may write
\be\label{CFTWishart2}
I(\boldsymbol{k})=\frac{1}{\Gamma(\boldsymbol \beta)}\int_0^\infty dx \exp \left( -\frac{x}{2\omega}\boldsymbol{k}^\top\Sigma_0\boldsymbol{k}\right)G_{ 0,N } ^{ N,0 }  \left( 
\begin{array}{c}
{-} \\ 
{ \boldsymbol\beta-{\bf 1}}
\end{array}
\bigg |x   \right),
\ee
where    the  $G$-function of one-variable now appears in the integrand.
Inserting (\ref{CFTWishart2}) into (\ref{int2}) and  rearranging terms yield
\begin{align}
P_N(\boldsymbol{r})&=\frac{1}{\Gamma(\boldsymbol{\beta})} \int_0^\infty dx \; G_{ 0,N } ^{ N,0 }  \left( 
\begin{array}{c}
{-} \\ 
{ \boldsymbol\beta-{\bf 1}}
\end{array}
\bigg |x   \right) \nonumber \cr
& \times \int \frac{d\boldsymbol{k}}{(2\pi)^p}e^{-i\boldsymbol{k}\cdot\boldsymbol{r}} \exp \left( -\frac{x}{2\omega}\boldsymbol{k}^\top\Sigma_0\boldsymbol{k}\right).
\end{align}
Now, comparing the second integral  above with (\ref{HS}) we can write
\be
\label{PNx}
P_N(\boldsymbol{r})=\frac{1}{\Gamma(\boldsymbol \beta)}\int_0^\infty dx P\left(\boldsymbol{r}\bigg |\frac{x}{\omega}\Sigma_0\right) G_{ 0,N } ^{ N,0 }  \left( 
\begin{array}{c}
{-} \\ 
{ \boldsymbol\beta-{\bf 1}}
\end{array}
\bigg |x   \right).
\ee
It is interesting to note that by virtue of the CFT transformation we have been able to express $P_N(\boldsymbol{r})$ as a simple compound distribution (rather than a matrix compounding), very much like in the univariate case; see (\ref{signal}).

Now using (\ref{Gauss2}) into (\ref{PNx}), we obtain
\ba
\label{PNx2}
P_N(\boldsymbol{r})
=&
\frac{\omega^{p/2}}{|2\pi \Sigma_0|^{1/2}\Gamma(\boldsymbol \beta)}\displaystyle \int_0^\infty dx \;G_{ 0,N } ^{ N,0 }  \left( 
\begin{array}{c}
{-} \\ 
{ \boldsymbol\beta-{\bf 1}}
\end{array}
\bigg |x   \right)  \\   & \hspace{-1.7cm}  \times   \ x^{-p/2}\exp\left(-\frac{\omega}{2x} {\bf r}^\top \Sigma^{-1}_0 {\bf r}\right) .\nonumber
\ea

As before, this integral can be explicitly performed using the integration formula for the product of two $G$-functions:
\be 
P_N(\boldsymbol{r})=
\frac{\omega^{p/2}}{{|2\pi\Sigma_0|^{1/2}}\Gamma(\boldsymbol\beta)}G_{0,N+1}^{N+1,0}\left( 
\begin{array}{c}
- \\ 
{ \boldsymbol\beta-\frac{p}{2}{\bf 1}},0
\end{array}
\bigg |\frac{\omega}{2}\boldsymbol{r}^\top\Sigma_0^{-1}\boldsymbol{r}\right).
\label{eq:PN3}
\ee 

\subsubsection{Inverse Wishart Class}

Another class of background distributions  $f_N(\Sigma_N)$ can be defined 
by choosing $f(\Sigma_i | \Sigma_{i-1})$ as an inverse Wishart distribution:
\ba \label{inverseWishart2}
    f(\Sigma_i | \Sigma_{i-1}) &= \frac{|\beta_i \Sigma_{i-1}|^{\beta_i+(p+1)/2} }{\Gamma_p (\beta_i + \frac{p+1}{2})} |\Sigma_i|^{-\beta_i-p-1}  \nonumber \\ \times& \exp{\left[-\beta_i \text{Tr}(\Sigma_i^{-1} \Sigma_{i-1})\right]},
\ea
where again $\beta_i>0$.
Now, using the matrix version of the Mellin transform (\ref{MTransformMatrix}) and using the  notation already defined in (\ref{Matrix-Mellin-G}), we obtain
\begin{equation}
    \label{inverse}
    f_N(\Sigma_N) = \frac{|\omega \Sigma_0|^{-(p+1)/2}}{\Gamma_p( \boldsymbol{\beta} + \frac{p+1}{2}{\bf 1})} \overline{G}^{0, N}_{N, 0} \left( 
    \begin{array}{c}
   -\boldsymbol{\beta} - \frac{p+1}{2}{\bf 1}\\
    -
    \end{array} \middle \vert \frac{\Sigma_N}{\omega \Sigma_0}
    \right).
\end{equation}
Inserting (\ref{inverse}) into (\ref{signal2}) we obtain
\ba
    P_N({\bf r}) &=& \frac{|\omega \Sigma_0|^{-(p+1)/2}}{\Gamma_p( \boldsymbol{\beta} + \frac{p+1}{2}{\bf 1})} \int d\Sigma_N P({\bf r} | \Sigma_N)  \nonumber \\ && \times \; \overline{G}^{0, N}_{N, 0} \left( 
    \begin{array}{c}
   -\boldsymbol{\beta} - \frac{p+1}{2}{\bf 1}\\
    -
    \end{array} \middle \vert \frac{\Sigma_N}{\omega \Sigma_0}
    \right).
\ea
By making $X = (\omega \Sigma_0)^{-1/2} \Sigma_N (\omega \Sigma_0)^{-1/2}$ and the invariance of Meijer $\bar{G}$-function of matrix argument we obtain
\ba \label{CFTInverse1}
P_N({\bf r}) 
    =&\frac{|2\pi\omega\Sigma_0|^{-1/2}}{\Gamma_p( \boldsymbol{\beta} + \frac{p+1}{2}{\bf 1})}  \int dX |X|^{-1/2} \exp\left(-\frac{{\bf r}^\top \Sigma_0^{-1/2} X^{-1} \Sigma_0^{-1/2} {\bf r}}{2\omega}\right) \nonumber \\ &\times \; \overline{G}^{0, N}_{N, 0} \left( 
    \begin{array}{c}
 \!\!\!  -\boldsymbol{\beta} - \frac{p+1}{2}{\bf 1}\\ \nonumber
    -
    \end{array} \middle \vert X 
    \right) \\    
\ea
Using the CFT transformation [see Eq.~(\ref{Meijer-Inverse Wishart CFT}) in Appendix \ref{Meijer-Wishart CFT subsection}], we get
\ba
\label{CFTInverse2}
    P_N({\bf r}) &= \frac{|2\pi\omega\Sigma_0|^{-1/2}}{\Gamma({\bf \boldsymbol{\beta^\prime}} + {\bf \boldsymbol{1}})}  \int dx \; x^{-1/2} \exp\left(-\frac{{\bf r}^\top \Sigma_0^{-1} {\bf r}}{2x\omega}\right) \nonumber\\ &\times \; G^{0, N}_{N, 0} \left( 
    \begin{array}{c}
   -\boldsymbol{\beta} - \frac{p+1}{2}{\bf 1}\\
    -
    \end{array} \middle \vert x 
    \right) \nonumber\\
    &= \frac{1}{\Gamma({\bf \boldsymbol{\beta}} + {\bf \boldsymbol{1}})} \displaystyle \int dx \; x^{(p-1)/2} P\left({\bf r}| x\omega\Sigma_0 \right) \nonumber\\ &\times \; G^{0, N}_{N, 0} \left( 
    \begin{array}{c}
   -\boldsymbol{\beta} - \frac{p+1}{2}{\bf 1}\\
    -
    \end{array} \middle \vert x 
    \right) 
\ea
The integral is easily done by the same procedure as in the Wishart class. 
We then find 
\begin{equation}
\label{inverse_wishart}
    P_N({\bf r}) = \frac{| 2 \pi \Sigma_0|^{-1/2}}{\omega^{p/2}\Gamma({\bf \boldsymbol{\beta}} + {\bf \boldsymbol{1}})} {G}^{1, N}_{N, 1} \left( 
    \begin{array}{c}
   -\boldsymbol{\beta} - \frac{p}{2}{\bf 1}\\
    0
    \end{array} \middle \vert \frac{{\bf r}^\top \Sigma_0^{-1} {\bf r}}{2\omega}
    \right).
\end{equation}

\section{One-dimensional Projection of Distributions} 

\label{sectionunidimensional}

As we are going to discuss applications of the matrix H theory formalism to return series from the  S\&P500 index, it is  convenient to project the theoretical multivariate distributions,   $P_N({\bf r})$,  into corresponding univariate versions that capture the main features of the full joint distribution. 

In other words, we wish to integrate $P_N({\bf r})$ over all but one variable $r_i=\tilde{r}$, so as to obtain a univariate distribution $P_N(\tilde{r})$, which is then easier to analyze and apply to the data.
To do so, we shall assume  that (over large time separations and upon proper normalization) the components of multivariate process ${\bf r}(t)$ are all comparable and statistically indistinguishable from each other. This implies that the large scale covariance matrix $\Sigma_0$ matrix can be taken as a multiple of the identity matrix: $\Sigma_0 = \epsilon_0 \mathbb{1} $, where $\epsilon_0$ is common variance of the series. 

Next, we note that under the assumption above  both Eqs.~(\ref{PNx2}) and (\ref{CFTInverse2}), upon making $x = \epsilon \omega/ \epsilon_0$ and $x= \epsilon/ \epsilon_0 \omega $, respectively,  assume the following form:
\begin{equation}
    P_N({\bf r}) = \int_0^{\infty} d\epsilon (2 \pi \epsilon)^{-p/2} \exp (-{\bf r}^\top {\bf r}/2\epsilon) f_N(\epsilon),
    \label{eq:compound}
\end{equation}
{where the background distribution $f_N(\epsilon)$ turns out to be given  by the same distributions of the univariate case, namely Eqs.~(\ref{gamma}) and (\ref{meijer1}) for the Wishart and inverse-Wishart classes, respectively.} Integrating (\ref{eq:compound}) over all but one of the variables $r_i=\tilde{r}$, we  obtain
\begin{equation} \label{univariateprojection}
    P_N(\tilde{r}) = \int_0^{\infty} d\epsilon \frac{1}{\sqrt{2\pi \epsilon}}\exp(-\tilde{r}^2/2\epsilon) f_N(\epsilon).
\end{equation}
Inserting (\ref{gamma}) into (\ref{univariateprojection}) yields for the Wishart class:
\be\label{wishartaggregated}
P_N(\tilde{r})=
\frac{\omega^{1/2}}{\sqrt{2\pi \epsilon_0}\Gamma(\boldsymbol\beta)}G_{0,N+1}^{N+1,0}\left( 
\begin{array}{c}
- \\ 
{ \boldsymbol\beta-{\bf 1/2}},0
\end{array}
\bigg |\frac{\omega \tilde{r}^2}{2\epsilon_0}\right).
\ee
Similarly for the inverse-Wishart class, using (\ref{meijer1}) in (\ref{univariateprojection}), we have
\be\label{inversewishartaggregated}
    P_N(\tilde{r})=
\frac{(2\pi \omega 
\epsilon_0)^{-1/2}}{\Gamma({\bf \boldsymbol{\beta}} + {\bf \boldsymbol{1}})}G_{N,1}^{1,N}\left( 
\begin{array}{c}
{ -\boldsymbol{\beta}- {\bf 1/2}} \\ 
0
\end{array}
\bigg |\frac{\tilde{r}^2}{2\omega \epsilon_0}\right).
\ee

{It is remarkable that the one-dimensional projected distributions, $P_N(\tilde{r})$, 
of the multivariate distribution for both classes are  the same as their respective univariate H theory distribution; compare (\ref{wishartaggregated}) with (\ref{eq:PN2})   and (\ref{inversewishartaggregated})  with (\ref{eq:PN1}).}

{Once we have a univariate distribution, we can then apply it to an aggregated version of the multivariate data ${\bf r}(t)$, as we discuss next. First we compute the  normalized time series (returns):
\be
    M_i(t) = \frac{r_i(t) - \langle r_i (t) \rangle}{\sigma_i},
\label{eq:Mi}
\ee
where $\langle r_i(t) \rangle$ and $\sigma_i$ are the average and standard deviation of the $i$-th  time series, respectively. Let us now group these $p$ normalized time series as a rectangular matrix, $M$, such that  $M_{ij}=r_i(j)$, with $i=1,...,p$ and $j=1,...,T$, where $T$ is the total length of each time series. This matrix is convenient to define the empirical correlation matrix, $C$, among the normalized returns:
\be
    C = \langle M M^\top \rangle / T.
\ee
}

{
The assets in the multivariate time series grouped in $M$ have, in principle, different statistical properties and cannot be directly compared. To overcome this difficulty we employ the concept of aggregated distribution of returns \cite{schmitt2013non, manolakis2023analysis}. The main idea here is to rewrite the multivariate process in a basis where all assets are normalized and there is no correlation among them. Since the true correlations of the underlying stochastic process are not known {\it a priori}, the best that can be done is to rewrite the process in a basis which diagonalizes the empirical correlation matrix $C$. 
We thus define a vector random variable $\bar{\bf r}$ by the following relation: $\overline{ \bf r}= U^\top {\bf r}$,
where the matrix $U$ diagonalizes $C^{-1}$, that is,
\be
    {\bf r}^\top C^{-1} {\bf r} 
    = {\bf r}^\top U \Lambda^{-1} U^\top {\bf r} 
    =
     \overline{{\bf r}}^\top \Lambda^{-1} \overline{ \bf r},
\ee
where $\Lambda^{-1}$ is a diagonal matrix (namely, $C^{-1}$ in its basis of eigenvalues). One additional step is to divide each process by their respective square-root eigenvalue, $\sqrt{\lambda_i}$, so as to normalize all processes to the same standard deviation, thus rendering them  comparable between one another. 
We thus define the normalized, uncorrelated processes  
 \be
    \tilde{\bf r} \equiv \Lambda^{-1/2} \, \overline{\bf r}.
\label{eq:tilder}
 \ee
}
{Considering the process $\tilde{\bf r}$ is useful because it makes the multiple time series essentially indistinguishable and uncorrelated, thus justifying the choice of  $\Sigma_0$ as a multiple of the identity matrix. Furthermore, since all processes are now comparable, we can aggregate the multiple time series, $\tilde{r}_i(t)$, $i=1,...,p$,  into a {\it single time series}, $R(t)$, $t=1,...,T$, which can then be analyzed in terms of the univariate projection distributions discussed above. In Sec.~\ref{applicationSection} we shall apply this formalism to financial data from the S\&P500 stock index.}

\section{One-Scale Background as a Particular Case}

In this section we discuss, for completeness,  the particular case  when the background dynamics has only one intermediate time scale, i.e., $N=1$, which recovers  previous results obtained within the context of superstatistics \cite{schmitt2013non, manolakis2023analysis}. 
In applications, however, one should avoid making  any {\it a priori} assumption about the specific value of $N$ and instead should  seek to determine it from the data, as it will be explained later.

\subsubsection{One-Scale Wishart Class}

The probability distribution $P_1(\tilde{r})$ can be written from (\ref{wishartaggregated}) as    
\begin{equation}
       P_1(\tilde{r}) = \frac{\beta^{1/2}}{\sqrt{2\pi \epsilon_0}\Gamma( \beta)} 
       G^{2, 0}_{0, 2} \left( 
    \begin{array}{c}
   -\\
    \beta - \frac{1}{2}, 0
    \end{array} \middle \vert \frac{\beta \tilde{r}^2}{2\epsilon_0}
    \right).
\end{equation}
Using the identity \cite{mathai2006generalized}
\begin{equation}
    {G}^{2, 0}_{0, 2} \left( 
    \begin{array}{c}
   -\\
    \alpha, \beta
    \end{array} \middle \vert z
    \right) = 2\pi^{\frac{\alpha + \beta}{2}}K_{\alpha - \beta} (2\sqrt{z}), 
\end{equation}
where $K_\nu(z)$ is the modified Bessel function of second kind,  we obtain:
\begin{equation}
    P_1(\tilde{r}) = \frac{2}{\sqrt{2}^\beta 
    \Gamma(\beta)}   \sqrt{\frac{\beta}{\pi\epsilon_0}} \sqrt{\frac{\beta \tilde{r}^2}{\epsilon_0}}^{\beta - 1/2} \! K_{\beta - 1/2} 
    \left(\sqrt{\frac{2\beta \tilde{r}^2} {\epsilon_0}}\right),
\end{equation}
{which is the same distribution found in Refs.~\cite{schmitt2013non}
using the matrix version of superstatistics with the assumption that the background fluctuates over all matrices in the Wishart ensemble.}

\subsubsection{One-Scale Inverse Wishart Class}

In this case, $P_1(r)$ can be written from (\ref{inversewishartaggregated}) as
\begin{equation}
\label{iw1}
    P_1(\tilde{r}) = \frac{(2\pi\beta \epsilon_0)^{-1/2}}{\Gamma(\beta + 1)} G^{1, 1}_{1, 1} \left( 
    \begin{array}{c}
   -\beta - \frac{1}{2}\\
    0
    \end{array} \middle \vert \frac{\tilde{r}^2}{2\beta \epsilon_0}
    \right).
\end{equation}

Using the property \cite{mathai2006generalized}
\begin{equation}
     \frac{1}{\Gamma(a)}{G}^{1, 1}_{1, 1} \left( 
    \begin{array}{c}
   1-a\\
    0
    \end{array} \middle \vert -z
    \right) = (1 - z)^{-a},
\end{equation}
we conclude that

\be
    P_1(\tilde{r}) = \frac{1}{\sqrt{2\pi \beta \epsilon_0}} \frac{\Gamma(\beta + 3/2)}{\Gamma(\beta + 1)} \left(1 + \frac{\tilde{r}^2}{2\beta \epsilon_0}\right)^{-\beta - 3/2}
    \label{eq:P1InvW}
\ee
{Again, the matrix H-theory formalism for $N=1$ reproduces previous results 
\cite{manolakis2023analysis}, where  distribution (\ref{eq:P1InvW}) is obtained via the matrix version of superstatistics assuming the background 
matrices are from the inverse Wishart ensemble.}
We emphasize, however, that the general family of distributions for arbitrary $N>1$ has not been investigated before in the context of multivariate time series.

\section{Application to stock markets} 

\label{applicationSection}

The data used here comprise 14 years (2010-2024) of daily closing quotes of 437 stocks listed on the S\&P500 index up to March, 6, 2024, {corresponding to 3565 data points for each stock}. A list of the stocks used in this work is available at \cite{SP500Changes2018}. Only stocks that were part of the index over the entire period above were considered in our analysis. (Owing to merges, acquisitions, and market capitalization changes some companies may have entered or left the S\&P500  during this  period \cite{SP500Changes} --- these stocks were removed from our data set.)

{Let us  consider the set of $p$ stock price series, $m_i(t)$, where $i=1,...,p$ and  $t=1,...,T$, with $p=437$ and $T=3565$.}
Here we shall analyze the asset logarithmic returns defined by
\be
r_i(t) = \ln(m_i (t + \Delta t)) - \ln(m_i(t)),
\ee
where $\Delta t=1$ day for smoothing purposes.  
 \be
    \tilde{\bf r} \equiv \Lambda^{-1/2} \, \overline{\bf r}.
 \ee 
 
\begin{figure}
     \centering
     \includegraphics[scale=0.38]{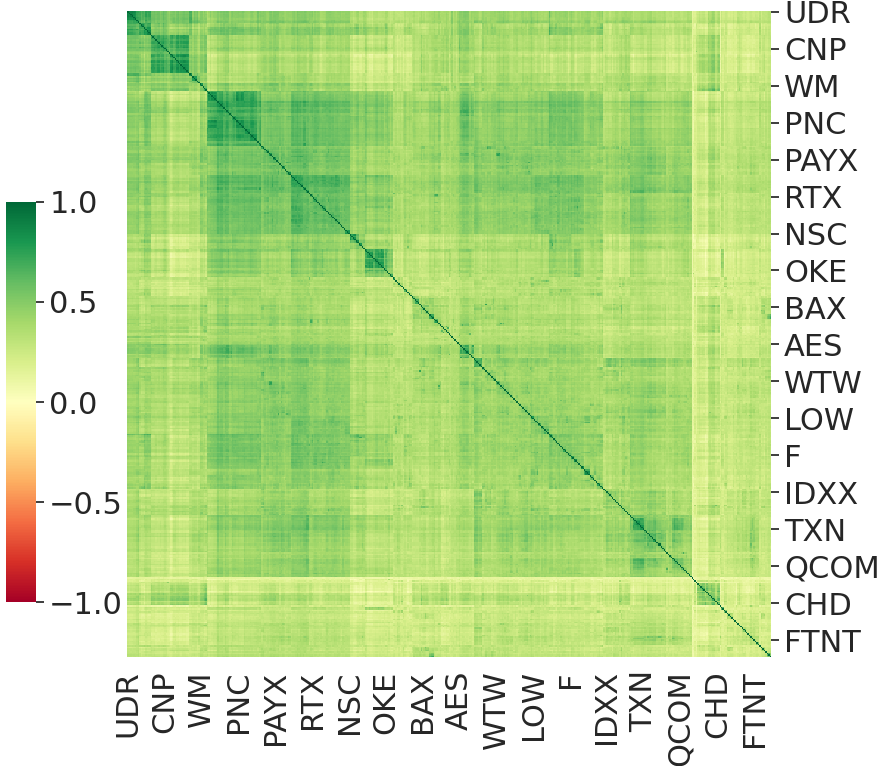}
     \caption{Correlation matrix for the normalized daily returns over a period of 14 years represented as a heat map. Stocks are sorted by clusters, indicating the existence of industrial branches in the S\&P 500 data (see \cite{mantegna1999introduction} for details on the covariance distance clustering method), where one sees strong correlations among stocks in a given cluster (green squares along the diagonal)  as well as correlations between different clusters (off diagonal rectangular structures). The overall behavior of the correlation matrix reveals the multivariate nature of the underlying stochastic process.}
     \label{fig:covariance matrix}
 \end{figure}

{The correlation matrix of the daily returns of the S\&P500  index is shown  in Fig.~\ref{fig:covariance matrix}. It is possible to see that in fact we have correlation among the assets, indicating that our assumption about the multivariate nature of process is verified, being very different from a identity matrix (indicating independent processes). Also the matrix is organized by clusters, indicating stronger correlations for groups of assets which is an evidence of the industrial branch arrangement already related in \cite{guhr2003new}.}

\begin{figure}
     \centering
     \includegraphics[scale=0.53]{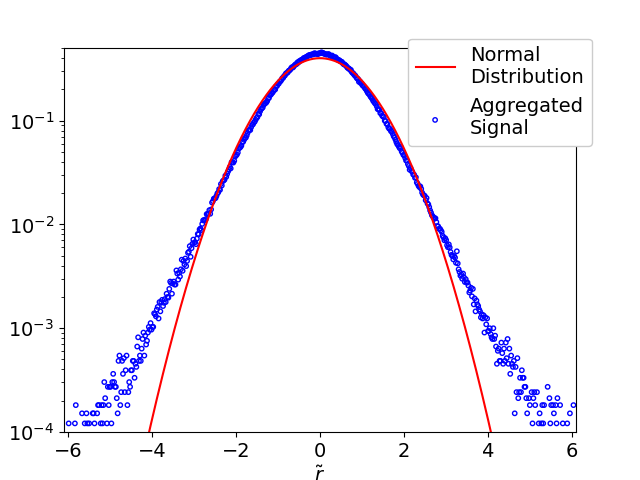}
     \caption{{Distribution of aggregated returns (blue open circles) superimposed with a Gaussian distribution (red solid line) of zero mean and unity variance. The non-Gaussian behavior of the  data  is clearly seen in the form of  heavy tails, i.e., a slower-than-Gaussian decay.}}
     \label{fig:aggregated histograms}
 \end{figure}

\begin{figure}
     \centering
     \includegraphics[scale=0.53]{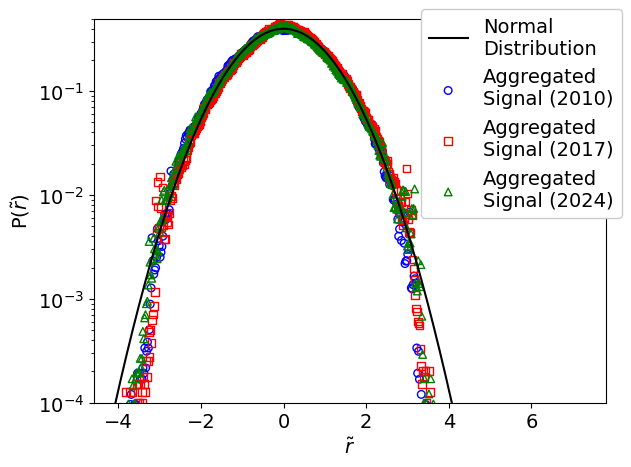}
     \caption{{Distributions of aggregated returns in a period of  10 days  in 2010 (blue open circles), 2017 (red open squares), and  2024 (green open triangles). Also show for comparison is a Gaussian distribution (black solid line), indicating that the returns at small time scales display Gaussian behavior.}}
     \label{fig:aggregated gaussianplot}
 \end{figure}

{As discussed in Sec.~\ref{sectionunidimensional}, after normalizing the vector $\mathbf{r}$ to zero mean and unit variance, see (\ref{eq:Mi}), and rotating it to the diagonal basis of the empirical covariance matrix, the  processes $\tilde{\bf r}$ defined in (\ref{eq:tilder}) become uncorrelated and statistically similar to one another, so that we can form a single, much larger time series, $R_i(t)$,  of aggregated returns with $437\times 3565= 1,557,905$ points.} 
Figure \ref{fig:aggregated histograms} shows the  probability distribution of the aggregated returns (blue circles), superimposed with a Gaussian distribution (red solid line) of zero mean and unit variance. [Recall that we are working with normalized returns, see (\ref{eq:Mi}).] From this figure one  sees that the empirical returns (over the entire period considered) display heavy tails that deviate substantially from  a Gaussian distribution. In comparison, Fig.~\ref{fig:aggregated gaussianplot} displays the aggregated distribution over a short period of time, namely 10 days in January of years 2010, 2017 and 2024. These limited (normalized) time series are  well described by a normal distribution (of zero mean and unit variance), in agreement with our hypothesis that the returns tend to be normally distributed over small periods of time. This further validates one of the tenets of H theory, namely that  the full distribution is to be obtained from the compounding of a Gaussian with a proper  background  distribution. Contrary to other compounding-based approaches \cite{schmitt2013non, manolakis2023analysis}, however, in the H theory formalism the background distribution is {\it not postulated a priori} but rather determined from the data, as described below.

In order to find the background empirical distribution for the aggregated returns, we first  extract a corresponding background time series from each individual process $\tilde r_i$. To do this, for each  $\tilde r_i(t)$ we compute an auxiliary series of  variance estimators over moving windows of size $L$: $\epsilon^L_i(t) = \frac{1}{L} \sum_{j=0}^{L-1} [\tilde{r}_i(t - j\delta t) - \langle \tilde{r}_i(t) \rangle_L]^2$, 
where $\langle \tilde{r}_i(t) \rangle_L = \frac{1}{L} \sum_{j=0}^{M-1} \tilde{r}_i(t - j \delta t)$. 

\begin{figure}
     \centering
     \includegraphics[width=1\linewidth]{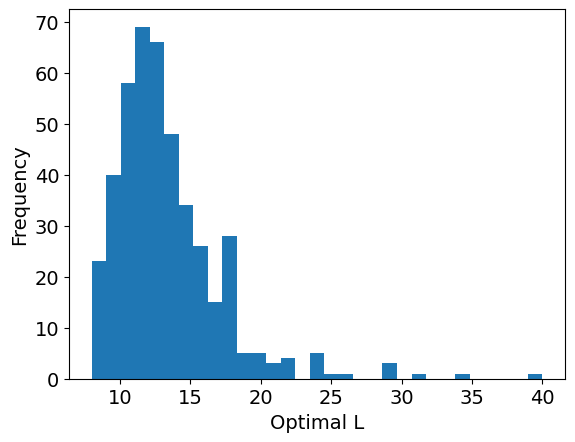}
     \caption{{Histogram of the optimal window sizes, $L$, for constructing the auxiliary variance series for  each return series $\tilde{r}_i(t)$. The mean value of this distribution is close to $L = 14$.}}
     \label{fig:M_hist}
 \end{figure}

To determine the optimal window size \( L \), we compound a normal distribution with the distribution of \(\epsilon^L_i(t)\) (see (\ref{univariateprojection})) and compare the resulting distribution with the empirical distribution of \(\tilde{r}_i\). The best \( L \) is chosen as the one that minimizes the corresponding Kullback-Leibler (KL) divergence. The distribution of the optimal \( L \) values obtained for each time series \(\tilde{r}_i(t)\) is presented in Fig.~\ref{fig:M_hist}. It is evident that most optimal \( L \) values are concentrated around the mean \( L = 14 \).

\begin{figure}
     \centering
     \includegraphics[width=1\linewidth]{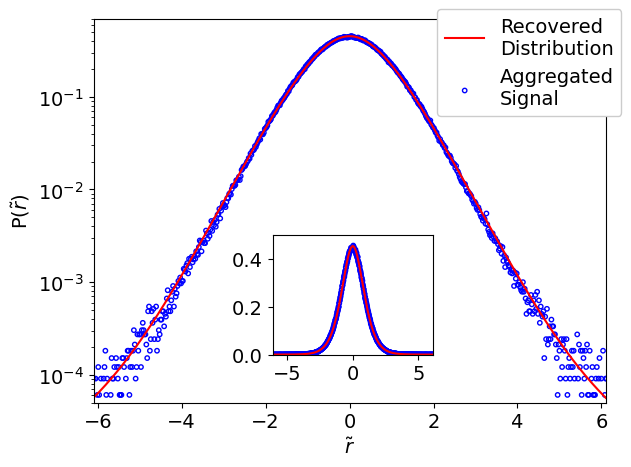}
     \caption{{Comparison between the recovered return distribution (red solid line), obtained by compounding a Gaussian with the  empirical distribution for the variances $\epsilon_i^L(t)$ with $L = 14$, and the empirical aggregated return distribution (blue open circles). The excellent agreement between the recovered and empirical distributions  confirms that the procedure for obtaining the aggregated background distribution explained in the text is consistent.}}
     \label{fig:recover}
 \end{figure}

{Next, we  construct a single variance series from the aggregated returns using a fixed window size, namely the mean value $L=14$, from which we then obtain the empirical background distribution associated with the full aggregated series. We verified that this procedure yields consistent results, in the sense that the  compounding of a Gaussian with the empirical variance distribution  does indeed recover the empirical distribution of aggregated returns, as indicated in Fig.~\ref{fig:recover}.}

\begin{figure}
     \centering
     \includegraphics[scale=0.52]{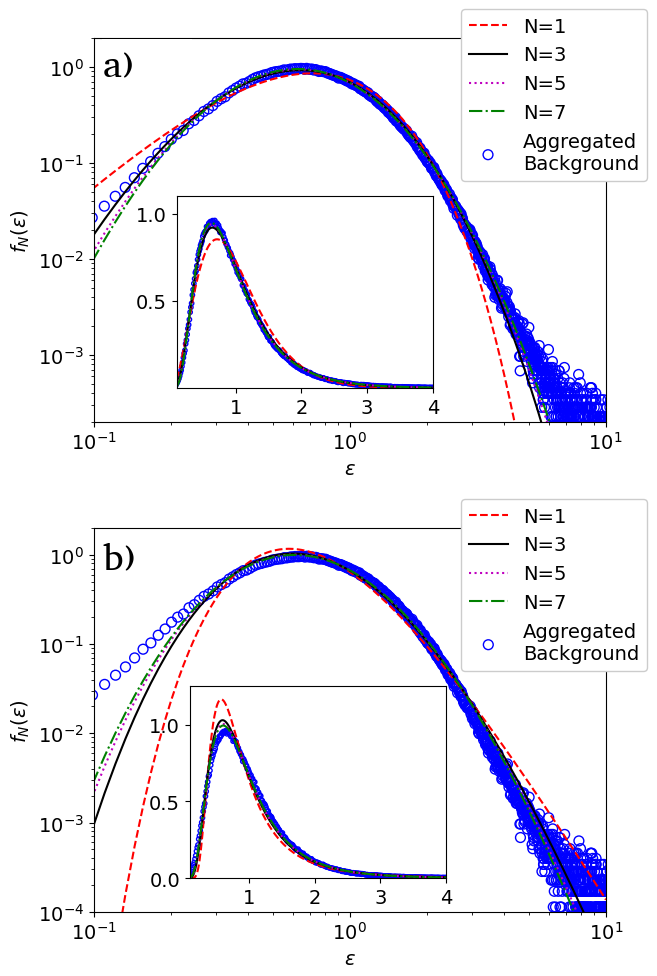}
     \caption{a) {Aggregated background distribution (open circles) and the best-fit theoretical background distributions, $f_N(\epsilon)$, for the  Wishart  class  and various values of $N$. The insert shows  the same curves as in the main panel but in bi-linear scale.  b) Same as in a) for the inverse Wishart  class.}}
     \label{fig:inverse wishart fit}
 \end{figure}

Having obtained  the empirical aggregated background distribution,  we can now perform a fitting procedure with the theoretical background distributions, $f_N(\epsilon)$, for both hierarchical models discussed in Sec.~\ref{sectionunidimensional}, namely Eqs.~(\ref{gamma}) and (\ref{meijer1}). 
In fitting these formulas for given $N$, we find the optimal value of the parameter $\beta$  that minimizes the KL Divergence between $f_N(\epsilon)$ and the aggregated background distribution. 
Figs.~\ref{fig:inverse wishart fit}(a) and \ref{fig:inverse wishart fit}(b) show the optimal fitting curves  for various values of $N$ for the Wishart and Inverse-Wishart classes, respectively, where the blue circles indicate the empirical background distributions and the solid curves, the theoretical fits. 
{Table \ref{tab:table} shows  the optimal values of $\beta$ for various $N$, where ones sees that  $\beta$   increases with $N$ --- a behavior that  can be explained as follows. As can be seen from Eqs.~(\ref{eq:asym2}) and (\ref{eq:asym1}), for fixed $N$ the tails of the theoretical curves $P_N(r)$ become less heavy as $\beta$ increases. Conversely, for fixed $\beta$ the  tails of  $P_N(r)$ become more elevated (i.e., heavier) as $N$ increases. Thus, when fitting a given data set (even when the fitting is performed at the background level), the value of $\beta$ must increase with $N$ to compensate the tendency of raising the tail for larger $N$.}

\begin{figure}
     \centering
     \includegraphics[scale=0.55]{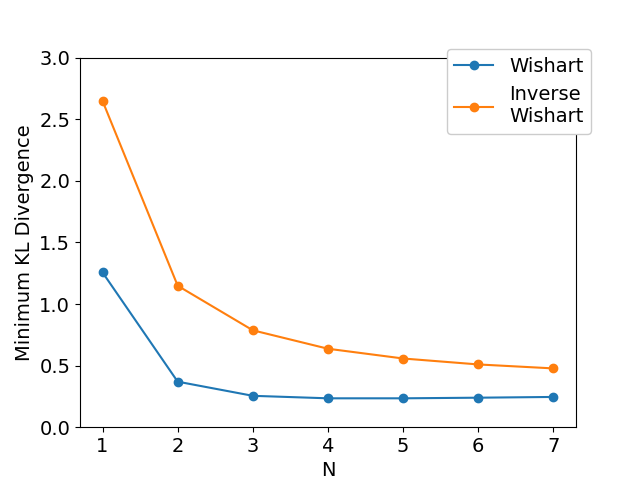}
     \caption{Error showed as the minimum KL Divergence for the optimal $\beta$ given a value of N.}
     \label{fig:errors fit}
 \end{figure}

Fig \ref{fig:errors fit} shows the fitting errors, defined as the minimum KL divergence for the optimal $\beta$, as a function of $N$ for both classes of models. In both cases the fitting errors decrease quickly as $N$ increases from $N=1$ to $N=3$ and tend to flatten out for $N>3$. This figures also shows that the 
Wishart class  provides a better description to the data, in the sense that it yields smaller fitting errors (for a given $N$) than the inverse-Wishart class. Furthermore, for the Wishart class there is hardly any improvement in the error fitting as $N$ increases past $N=3$. In fact, visual inspection of Fig.~\ref{fig:inverse wishart fit}(a) reveals that the left tail of the theoretical curve moves farther away from the data for $N>3$, with only a minimal improvement in the right tail. {Also, computation of the theoretical distributions is more costly for $N>3$ (because of the higher order $G$-functions) without  yielding any significant improvement to the agreement with the data. So we can  conclude that the best model that describes our data is the Wishart class with $N=3$ hierarchical levels.}

\begin{figure}
     \centering
     \includegraphics[scale=0.50]{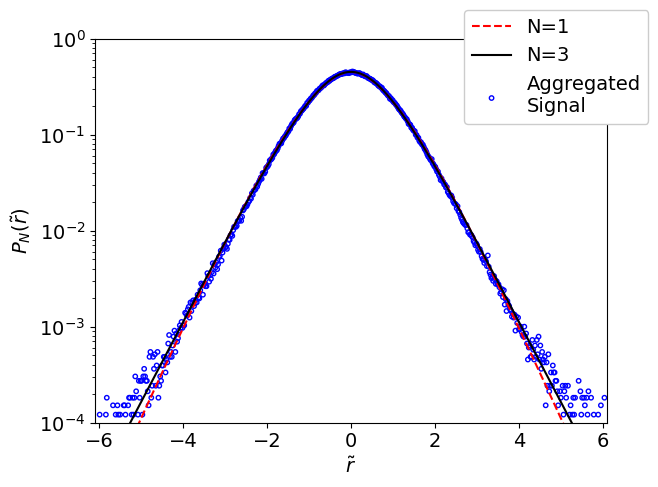}
     \caption{{Aggregated return distribution (open blue circles) and the theoretical distributions $P_N(\tilde{r})$ for the  Wishart  class   for $N = 1$ (dashed red line) and $N = 3$ (solid black line).}}
     \label{fig:wishart fit}
 \end{figure}

After having determined the best background model and its best-fit parameters, we can then plot the corresponding distribution, $P_N(r)$, for the aggregated returns, see Eq.~(\ref{wishartaggregated}), and compare the theoretical prediction with the empirical data. 
In Fig.~\ref{fig:wishart fit} we plot the theoretical distribution (solid black line) for the  Wishart class with $N=3$, which as shown above gives the best description to our data. Also shown in this figure, for comparison, is the best fit for  $N=1$, which yields a considerably poorer fit to the data, especially in the tails. We recall that the H theory with $N=1$ recovers the so-called superstatistics approach used, e.g., in Refs.~\cite{schmitt2013non,manolakis2023analysis}. Although the  theoretical prediction in this case ($N=1$) is also in somewhat good agreement with the return data, except at the tails, see Fig.~\ref{fig:wishart fit} (dashed red line), the agreement becomes much poorer when one looks at the background  distribution, see Fig.~\ref{fig:inverse wishart fit}(a).  In other words, by using a more stringent criterion, namely comparison at the background level, we can rule out $N=1$ as the best scenario for the data and instead conclude that at least $N=3$ relevant time scales are present in the underlying price dynamics. 

The presence of different time scales in financial markets is typically associated with the existence of investors with different investment horizons. In other words, traders with different investment horizons  have different investment strategies and consequently different response times to the market fluctuations,  which are then reflected in the underlying price dynamics \cite{vasconcelos2024turbulence}. Of course, it not easy to estimate {\it a priori} the number and size of relevant time scales in a given dataset for a given market. Using the analog of energy cascade in turbulence, which translates into an information cascade in financial markets \cite{vasconcelos2024turbulence}, one  expects that the intermediate time scales would be of the form $\tau_i=\tau_0/b^i$,  $i=1,..,N$, where $\tau_0$ is the largest relevant time scale and $b>1$ is some scaling factor (typically a small integer) proper of the corresponding cascade dynamics \cite{Schertzer1997}. Assuming $\tau_0$ to be of the order of a year ($\sim260$ business days), which is reasonable for our dataset, and choosing a sensible value for $b$, say $b=4$ \cite{Schertzer1997},
we  obtain that  the three intermediate time scales (from smaller to higher), $\tau_3$, $\tau_2$, and $\tau_1$, in our analysis ($N=3$) are of the order of one week, one month, and one quarter, respectively. These estimates seem  reasonable for  the U.S.~stock market as tracked by the S\&P500 index.
Recently, it has also been suggested (in the context of foreign exchange markets) that the number $N$ of relevant time scales should increase with market size \cite{vasconcelos2024turbulence}. It is thus an interesting question whether the same behavior is observed for stock indexes --- a problem that we plan to investigate in the future using the H-theory formalism.

 \begin{table}[]
\centering
 \renewcommand{\arraystretch}{1.2}
\begin{tabular}{|c|c|c|c|c|}
    \hline
   \multirow{ 2}{*}{N} & \multicolumn{2}{c|}{Inverse Wishart} & \multicolumn{2}{c|}{Wishart} \\ \cline{2-5} 
      & \(\beta\) & KL Div. & \(\beta\) & KL Div. \\ 
    \hline
    1 & 2.74 & 2.6479 & 3.49 & 1.2562  \\ 
    \hline
    2 &5.85 & 1.1481& 6.58 & 0.3695 \\ 
    \hline
    3 &8.95 & 0.7864& 9.67 & 0.2547 \\ 
    \hline
    4 & 12.05 & 0.6367 & 12.77 & 0.2344 \\ 
    \hline
    5 & 15.15 & 0.5575 & 15.87 & 0.2344  \\ 
    \hline
    6 &  18.25 & 0.5093  & 18.97 & 0.2394 \\ 
    \hline
    7 &  21.35 & 0.4772  & 22.07 & 0.2455 \\ 
    \hline
\end{tabular}
\caption{{Optimal fitting parameters found in Fig.~\ref{fig:inverse wishart fit} and corresponding errors (KL divergence).}}
\label{tab:table}
\end{table}

\section{Conclusion}

{The matrix extension of H theory  (MHT) presented in this work provides a systematic approach to extracting the number of scales from multivariate stochastic processes, categorizing them into two universality classes: the Wishart and the inverse Wishart classes. These two classes are generalizations of their univariate counterparts, namely the gamma and inverse gamma classes \cite{macedo2017universality},  respectively, thus extending the applicability of the theory to more complex systems. The MHT description of multivariate processes finds mathematical support in the generalization of Meijer $G$-functions to functions of matrix argument and in the application of CFTs theorems, justifying the use of aggregated distributions in the analysis. An examination of S\&P 500 stock returns provided compelling evidence for multiscale dynamics in financial markets when analyzing processes across multiple assets simultaneously, with the assumption that the covariance matrix evolves as a random matrix process. The identification of the Wishart universality class as the most appropriate description not only aligns with previous research but also offers a more precise characterization of the non-Gaussian behavior observed in financial markets over extended periods of time by estimating the number of characteristic scales in the underlying dynamics. Furthermore, this finding elucidates the origin of heavy tails as a consequence of interactions between multiple relevant scales.}

{The implications of this work covers various aspects of portfolio management. For instance, knowledge of the number of relevant scales in the time series could lead to more accurate descriptions of the tails of the probability density functions, potentially improving methods such as Value at Risk \cite{glasserman2002portfolio}. In addition, MHT may contribute to more precise estimations of structural credit risk by elucidating proper correlations between stocks, a concept previously explored using RMT in \cite{munnix2014random} and \cite{muhlbacher2018credit}. Additionally, portfolio optimization strategies may benefit from this insight, as the time horizon can influence portfolio composition when employing appropriate risk measures, as demonstrated in \cite{muzy2001multifractal}. Beyond its applications in econophysics, this theoretical framework may find utility in the description of diverse physical systems, including turbulence in fluids, random lasers and complex phenomena in condensed matter physics, opening avenues for interdisciplinary research and analysis.}

\section{Acknowledgements}  
This work was supported in part by the following Brazilian
agencies Conselho Nacional de Desenvolvimento Cient\'fico
e Tecnol\'ogico (CNPq), under Grants No. 608 307385/2023-0
(G.L.V.), No. 307626/2022-9 (A.M.S.M.), No. 303192/2022-
4, and No. 402519/2023-0 (R.O.); Ci\^encia e Tecnologia do
Estado da Bahia (FAPESB), Grant No. APP0021/2023 (R.O.);
and Coordena\c c\~ao de Aperfei\c coamento de Pessoal de N\'ivel Superior (CAPES), Grant No. 001.

\appendix 

\section{The Meijer $G$-function}\label{meijer G appendix}

The Meijer $G$-function is defined by its Mellin transform \cite{beals2013meijer}:
\be\label{Mtransform}
{\cal M}[G;s]=\frac{\prod_{j=1}^m\Gamma
(s+b_j)\prod_{j=1}^n\Gamma (1-s-a_j)}{\prod_{j=m+1}^q\Gamma
(1-s-b_j)\prod_{j=n+1}^p\Gamma (s+a_j)},
\ee
where
\be 
{\cal M}[f;s] \equiv \int_0^\infty dx  x^{s-1}f(x).
\ee 
The standard notation is
\be 
G(x)=G_{p,q}^{m,n}\left( 
\begin{array}{l}
\boldsymbol{a} \\ 
\boldsymbol{b}
\end{array}
\bigg | x\right) \label{Univariate-Mellin-G},
\ee 
where $\boldsymbol{a}=(a_1,\ldots ,a_p)$ and $\boldsymbol{b}=(b_1,\ldots ,b_q)$. {When one of the indexes $\{m,n,p,q\}$ is zero in Eq.~\eqref{Univariate-Mellin-G}, the corresponding product of gamma functions in the Mellin transform in (\ref{Mtransform}) is not present  and so  we  leave a blank slot, denoted by the symbol $-$ in the respective parameter entry of the $G$-function.}

A few useful properties of the Meijer $G$-functions, {which help to perform some of the calculations indicated in the main text},  are listed below.
\begin{itemize}
    \item {\it Argument inversion:}
    \be\label{argumentinversion}
    G_{p,q}^{m,n}\left( 
\begin{array}{l}
\boldsymbol{a} \\ 
\boldsymbol{b}
\end{array}
\bigg | \frac{1}{x}\right)= 
G_{q,p}^{n,m}\left( 
\begin{array}{l}
{\bf 1} - \boldsymbol{b} \\ 
{\bf 1} - \boldsymbol{a}
\end{array}
\bigg | x\right).
    \ee

\item {\it Power absorption:}
\be\label{powerabsorption}
x^{\sigma} G_{p,q}^{m,n}\left( 
\begin{array}{l}
\boldsymbol{a} \\ 
\boldsymbol{b}
\end{array}
\bigg | x\right)= 
G_{p,q}^{m,n}\left( 
\begin{array}{l}
\sigma{\bf 1} + \boldsymbol{a} \\ 
\sigma{\bf 1} + \boldsymbol{b}
\end{array}
\bigg | x\right),
\ee
\item {\it {Convolution theorem:}}
\ba\label{convolution}
    && \int_{0}^{\infty} 
    G_{n,m}^{m,n}\left( 
    \begin{array}{l}
        \boldsymbol{a} \\ 
        \boldsymbol{b}
    \end{array}
    \bigg | \xi x\right) 
    G_{t,r}^{r,t}\left( 
    \begin{array}{l}
        \boldsymbol{c} \\ 
        \boldsymbol{d}
    \end{array}
    \bigg | \eta x\right) dx \nonumber \\
    && \; = \frac{1}{\eta} G_{n + r,m + t}^{m + t,n + r}\left( 
    \begin{array}{l}
        (\boldsymbol{a}, -\boldsymbol{d}) \\ 
        (\boldsymbol{b}, -\boldsymbol{c})
    \end{array}
    \bigg | \frac{\xi}{\eta} \right). \label{convolution1}
\ea
\end{itemize}

{The last  property above is particularly useful when computing integrals involving $G$-functions, since it  shows that the integral of the product of two $G$-functions is also a $G$-function \cite{debnath2016integral}.}

\section{Properties of the Meijer $\bar G$-function of matrix argument}

\label{appendixproperties}

{The Meijer $\bar{G}$-function of matrix argument  was defined in  the main text in terms of its matrix Mellin transform, see Eqs.~(\ref{eq:MbarG}) and (\ref{MTransformMatrix}), which we reproduce here for convenience:
\be\label{M}
\begin{split}
M[\bar{G};s] = \ &\frac{\prod_{j=1}^m\Gamma_p
(s+b_j)}{\prod_{j=m+1}^q\Gamma_p
((1+p)/2-s-b_j)} \\ &\times\frac{\prod_{j=1}^n\Gamma_p ((1+p)/2-s-a_j)}{\prod_{j=n+1}^p\Gamma_p (s+a_j)}.
\end{split}
\ee
where 
\be 
M[f;s] \equiv \int_{X>0} dX |X|^{s-(p+1)/2}f(X),
\ee 
with $X$ being a $p\times p$ real, symmetric, positive definite matrix and $\Gamma_p$ is the multivariate version of the gamma function; see Eq.~(\ref{multivariategamma}).
}

{Here we prove two basic properties of the $\bar{G}$ function of matrix argument which are used in the main text.
\begin{itemize}
    \item {\bf Argument inversion}:
\end{itemize}
\be \label{matrixargumentinversion}
\bar{G}_{p,q}^{m,n}\left( 
\begin{array}{l}
\boldsymbol{a} \\ 
\boldsymbol{b}
\end{array}
\bigg | X^{-1}\right) =
\bar{G}_{q,p}^{n,m}\left( 
\begin{array}{l}
\frac{1+p}{2}\boldsymbol{1}-\boldsymbol{b} \\ 
\frac{1+p}{2}\boldsymbol{1}-\boldsymbol{a}
\end{array}
\bigg | X\right)
\ee 
{\it Proof:}
Starting from 
\be 
M[\bar{G};s] \equiv \int_{X>0} dX |X|^{s-(p+1)/2}\bar{G}(X),
\ee 
and making the change of variable $X = Y^{-1}$, so that $dX = |Y|^{-p-1} dY$, we obtain
\be 
M[\bar{G};s] \equiv \int_{Y>0} dY |Y|^{-s-(p+1)/2}\bar{G}(Y^{-1}).
\ee 
From the definition of (\ref{M}), we obtain after some manipulation:
\ba
M[\bar{G};s]=&\frac{\prod_{j=1}^m\Gamma_p
(s+b_j)}{\prod_{j=m+1}^q\Gamma_p
((1+p)/2-s-b_j)} \nonumber \\ 
&\times\frac{\prod_{j=1}^n\Gamma_p ((1+p)/2-s-a_j)}{\prod_{j=n+1}^p\Gamma_p (s+a_j)} \nonumber \\ 
=&\frac{\prod_{j=1}^n\Gamma_p (-s+(1+p)/2-a_j)}{\prod_{j=n+1}^p\Gamma_p (s+a_j)}\\ &\times \frac{\prod_{j=1}^m\Gamma_p
(s+b_j)}{\prod_{j=m+1}^q\Gamma_p
(-s+(1+p)/2-b_j)} \nonumber\\
=&\frac{\prod_{j=1}^n\Gamma_p (-s+(1+p)/2-a_j)}{\prod_{j=n+1}^p\Gamma_p ((1+p)/2+s+a_j-(1+p)/2)} \nonumber 
\\ &\times\frac{\prod_{j=1}^m\Gamma_p
((1+p)/2+ s+b_j - (1+p)/2)}{\prod_{j=m+1}^q\Gamma_p
(-s+(1+p)/2-b_j)} \nonumber.
\ea
Comparing the previous two equations, we  see that $\bar{G}(Y^{-1})$ satisfies the Mellin matrix transform for the variable $-s$, after making the correspondence $a_j \rightarrow (1+p)/2 - b_j$ and $b_j \rightarrow (1+p)/2 - a_j$ and $m \leftrightarrow n$, $q \leftrightarrow p$, from which  (\ref{matrixargumentinversion}) follows.
}

{
\begin{itemize}
    \item {\bf Power Absorption:}
\end{itemize}
\be \label{matrixpowerabsorption}
|X|^{\sigma}\bar{G}_{p,q}^{m,n}\left( 
\begin{array}{l}
\boldsymbol{a} \\ 
\boldsymbol{b}
\end{array}
\bigg | X\right) =
\bar{G}_{p,q}^{m,n}\left( 
\begin{array}{l}
\sigma\boldsymbol{1}+\boldsymbol{a} \\ 
\sigma\boldsymbol{1}+\boldsymbol{b}
\end{array}
\bigg | X\right)
\ee 
}
{
{\it Proof:}
The Mellin matrix transform of $|X|^{\sigma}G(X)$ can be worked out similarly using the definition of the matrix-argument Meijer $G$-function:
\ba
M[\bar{G};s] &= \int_{X>0} dX |X|^{s-(p+1)/2} |X|^{\sigma} \bar{G}(X)\cr
&= \int_{X>0} dX |X|^{s+\sigma-(p+1)/2} \bar{G}(X)
\ea 
The expression above is simply the Mellin transform on the variable $s + \sigma$, where $\sigma$ can be absorbed by $a_i$ and $b_j$ using the definition:
\be
\begin{split}
M[|X|^{\sigma} \bar{G}; s]=&\frac{\prod_{j=1}^m\Gamma_p
(s+(\sigma+b_j))}{\prod_{j=m+1}^q\Gamma_p
((1+p)/2-s-(\sigma+b_j))}\\ \times&\frac{\prod_{j=1}^n\Gamma_p ((1+p)/2-s-(\sigma+a_j))}{\prod_{j=n+1}^p\Gamma_p (s+\sigma+a_j)} 
\end{split}
\label{eq:MXsigma}
\ee
that is, letting $a_i \rightarrow a_i + \sigma$ and $b_i \rightarrow b_i + \sigma$. From (\ref{eq:MXsigma}) one then obtains (\ref{matrixpowerabsorption}).
}

\section{Color Flavor Transformation Proof}
\label{appendixA}
\subsection{Gamma-CFT}
Let us first start with the integral 
\ba \label{defGpIntegral}
    G_p(\nu, A, {\bf r})&\equiv\frac{1}{\Gamma_p(\nu)} \displaystyle \int_{X>0} dX |X|^{\nu - (p+1)/2} \nonumber
    \\ \times& \exp(-\text{Tr} X -{\bf r}^\top A^{1/2} X A^{1/2}{\bf r})
\ea
where $A$ and $X$ are p $\times$ p real symmetric positive definite matrices $\text{Re}(\nu) > (p+1)/2$. Writing ${\bf r}^\top A^{1/2} X A^{1/2}{\bf r} = \text{Tr}(A^{1/2} {\bf r}\;{\bf r}^\top A^{1/2} X)$ we obtain \cite{mathai1997jacobians}
\begin{equation}
    G_{p}(\nu, A, {\bf r}) = |\mathbb{1}_{p}+A^{1/2} {\bf r}\;{\bf r}^\top A^{1/2} |^{-\nu} ,
\end{equation}
where $\mathbb{1}_{p}$ is a $p \times p$ identity matrix.
Using the Weinstein–Aronszajn identity \cite{pozrikidis2014introduction},
\begin{equation}\label{defG1Integral}
\begin{split}
    G_{p}(\nu, A, {\bf r}) &= |\mathbb{1}_{p}+A^{1/2} {\bf r}\;(A^{1/2} {\bf r})^\top |^{-\nu} \\
    &= (1 + {\bf r}^\top A {\bf r} )^{-\nu} \\
    &= \frac{1}{\Gamma(\nu)}\int_0^\infty dx\; x^{\nu - 1} \exp(-x-x{\bf r}^\top A {\bf r})\\
    &= G_1(\nu, A, {\bf r})
\end{split}
\end{equation}

We can conclude that the value of $G_p$ is independent of the matrix dimension. It makes easier to substitute all integrals defined in Eq.~(\ref{defGpIntegral}) by their univariate versions $G_1$.

\subsection{Meijer-Wishart CFT} \label{Meijer-Wishart CFT subsection}
Let us consider in this section that $f(\Sigma_i|\Sigma_{i-1})$ is the Wishart distribution in Eq.~(\ref{wishartdistribution}). The Meijer $\bar{G}$-function with matrix argument can be written as:
\begin{equation}
\begin{split}
    &\frac{1}{\Gamma_p({\bf \boldsymbol{\beta}})} \overline{G}^{N, 0}_{0, N} \left( 
    \begin{array}{c}
    - \\
    \boldsymbol{\beta} - \frac{p+1}{2}{\bf 1}
    \end{array} \middle \vert X
    \right)\\& = \int d\Sigma_1 ... d\Sigma_{N-1} f(X|\Sigma_{N-1})...f(\Sigma_1|\mathbb{1} \omega)
\end{split}
\end{equation}
Let us start with the integral we want to work out
\begin{align}
    \label{Ik}
    \mathcal{M}G_{p, N}(A, \boldsymbol{\beta}, {\bf r}) &\equiv \nonumber \\\frac{1}{\Gamma_p({\bf \boldsymbol{\beta}})} \int &dX \exp(-{\bf r}^\top A^{1/2} X A^{1/2} {\bf r} )\\ \times &\overline{G}^{N, 0}_{0, N} \left( 
    \begin{array}{c}
    - \\
    \boldsymbol{\beta} - \frac{p+1}{2}{\bf 1}
    \end{array} \middle \vert X
    \right) \nonumber \\
    =\int d\Sigma_1 ... d\Sigma_{N-1} &f(\Sigma_{N-1}|\Sigma_{N-2})...f(\Sigma_1|\mathbb{1}\omega) \label{MGcompound}\\
\times \int dX &\exp (- {\bf r}^\top A^{1/2} X A^{1/2} {\bf r}) \; f(X|\Sigma_{N-1}).\nonumber
\end{align}
The integral in $dX$ in last equality above can be developed using 
\begin{equation}
\begin{split}
  &\int dX \exp (- {\bf r}^\top A^{1/2} X A^{1/2} {\bf r}) \; f(X|\Sigma_{N-1}) \\
  = &\frac{|\beta_N \Sigma_{N-1}^{-1}|^{\beta_N}}{\Gamma(\beta_N)} \int dX |X|^{\beta_N - (p+1)/2} \\ &\times \exp(-\beta_N\text{Tr}(X \Sigma_{N-1}^{-1})-{\bf r}^\top A^{1/2} X A^{1/2} {\bf r})
\end{split}
\end{equation}
performing a change of variables $X = \Sigma_{N-1}^{1/2} Y \Sigma_{N-1}^{1/2}/\beta_N$, we can use the Gamma-CFT in Eq.~(\ref{defG1Integral}),
\begin{equation*}
    \begin{split}
         = &\frac{1}{{\Gamma_p(\beta_N)}}\int dY |Y|^{\beta_N - (p+1)/2} \\ & \times \exp(-\text{Tr}Y- {\bf r}^\top A^{1/2} \Sigma_{N-1}^{1/2} Y \Sigma_{N-1}^{1/2} A^{1/2} {\bf r}/\beta_N) \\
         = &\frac{1}{{\Gamma(\beta_N)}} \int dy\; y^{\beta_N - 1} \\
         &\times \exp(-y -y{\bf r}^\top A^{1/2} \Sigma_{N-1} A^{1/2}{\bf r}/\beta_N ) \\
         = &\frac{\beta_N^{\beta_N}}{{\Gamma(\beta_N)}}\int dx\; x^{\beta_N - 1} \\
         &\times \exp(-x\beta_N  -x{\bf r}^\top A^{1/2} \Sigma_{N-1} A^{1/2}{\bf r} ) ,
    \end{split}
\end{equation*}
where in the last equality we made $y = x\beta_N$. Putting the result above back in Eq.~(\ref{MGcompound}) and rearranging the integral ordering, we obtain
\begin{equation}
\begin{split}
    &\mathcal{M}G_{p, N}(A, \boldsymbol{\beta}, {\bf r}) = \\
    &\frac{\beta_N^{\beta_N}}{{\Gamma(\beta_N)}}\int dx\; x^{\beta_N - 1} \exp(-x\beta_N) \\
    \times&\int d\Sigma_1 ... d\Sigma_{N-2} f(\Sigma_{N-2}|\Sigma_{N-3})...f(\Sigma_1|\mathbb{1}\omega) \\
\times & \int d\Sigma_{N-1}\exp(-{\bf r}^\top A^{1/2} \Sigma_{N-1} A^{1/2}{\bf r}/x ) \; f(\Sigma_{N-1}|\Sigma_{N-2}).\nonumber
\end{split}
\end{equation}

The whole expression above has the same form of integral $dX$ as in Eq.~(\ref{Ik}). Repeating the procedure described above until we run out of matrix integrals we obtain
\begin{equation}
    \begin{split}
        &\mathcal{M}G_{p, N}(B, \boldsymbol{\beta}, {\bf r}) = \frac{\beta_N^{\beta_N} ... \beta_1^{\beta_1}}{{\Gamma\left(\boldsymbol{\beta}\right)}}\\ &\times \int dxx^{\beta_N - 1}  
     \exp(-x\beta_N) \prod_{i=1}^{N-1} \int d\varepsilon_{i}\varepsilon_{i}^{\beta_i - 1}  
     \exp(-\varepsilon_{i} \beta_i) \\ 
     &\times \exp(-\omega{\bf r}^\top A {\bf r}x\varepsilon_{N-1}...\varepsilon_1)
    \end{split}
\end{equation}
After performing the change of variables $\varepsilon_i = \epsilon_{i}/\epsilon_{i-1}$, where $\varepsilon_N = x$ and $\epsilon_0 = \omega$ and rearranging the integrals in $d\epsilon_i$ we obtain the Meijer $G$-function with scalar arguments.
\begin{equation} \label{Meijer-Wishart CFT}
    \begin{split}
        \mathcal{M}G_{p, N} (A, \boldsymbol{\beta}, {\bf r}) &=
        \frac{1}{\Gamma(\boldsymbol{\beta})}
        \int dx \exp(-x{\bf r}^\top A {\bf r}) \\&\times \; G^{N, 0}_{0, N} \left( 
    \begin{array}{c}
   -\\
    \boldsymbol{\beta} - {\bf 1}
    \end{array} \middle \vert x 
    \right) = \mathcal{M} G_{1, N} (A, \boldsymbol{\beta}, {\bf r})
    \end{split}
\end{equation}
In Eqs.~(\ref{CFTWishart1}) and (\ref{CFTWishart2}), it is enough to make $B = \Sigma_0\sqrt{2\omega}$.

From Eq.~(\ref{Meijer-Wishart CFT}), making a changing of variables $X = Y^{-1}$, so that $dX = |Y|^{-p-1} dY$ in the multivariate integral and $x = y^{-1}$, so that $dx = -y^{-2} dy$ and making $\boldsymbol{\beta} \rightarrow \boldsymbol{\beta} + \frac{3}{2} \boldsymbol{1}$ we obtain
\begin{align}
    \frac{1}{\Gamma_p({\bf \boldsymbol{\beta}} + \frac{3}{2}\boldsymbol{1})} \int &dY |Y|^{-p-1} \exp(-{\bf r}^\top A^{1/2} Y^{-1} A^{1/2} {\bf r} ) \nonumber \\ \times &\; \overline{G}^{N, 0}_{0, N} \left( 
    \begin{array}{c}
    - \\
    {\bf \boldsymbol{\beta}} + \frac{3}{2}\boldsymbol{1} - \frac{p+1}{2}{\bf 1}
    \end{array} \middle \vert Y^{-1}
    \right) \nonumber \\ =\frac{1}{\Gamma({\bf \boldsymbol{\beta}} + \frac{3}{2}\boldsymbol{1})}
        \int &dy y^{-2} \exp(-{\bf r}^\top A {\bf r}/y) \nonumber \\ \times &\; G^{N, 0}_{0, N} \left( 
    \begin{array}{c}
   - \\
    {\bf \boldsymbol{\beta}} + \frac{3}{2}\boldsymbol{1} - {\bf 1}
    \end{array} \middle \vert y^{-1} 
    \right) \nonumber
\end{align}
By using argument inversion in Eqs.~(\ref{argumentinversion}) and  (\ref{matrixargumentinversion}), we obtain
\begin{align}
    \frac{1}{\Gamma_p({\bf \boldsymbol{\beta}} + \frac{3}{2}\boldsymbol{1})} \int &dY |Y|^{-p-1} \exp(-{\bf r}^\top A^{1/2} Y^{-1} A^{1/2} {\bf r} ) \nonumber \\ \times &\; \overline{G}^{0, N}_{N, 0} \left( 
    \begin{array}{c}
    -{\bf \boldsymbol{\beta}} - \frac{3}{2}\boldsymbol{1} + (p+1){\bf 1} \\
    -
    \end{array} \middle \vert Y
    \right) \nonumber \\ =\frac{1}{\Gamma({\bf \boldsymbol{\beta}} + \frac{3}{2}\boldsymbol{1})}
        \int &dy y^{-2} \exp(-{\bf r}^\top A {\bf r}/y) \nonumber \\ \times &\; G^{0, N}_{N, 0} \left( 
    \begin{array}{c}
   -{\bf \boldsymbol{\beta}} + \frac{1}{2}\boldsymbol{1}\\
    -
    \end{array} \middle \vert y 
    \right) \nonumber
\end{align}
and using power absorption in Eqs.~(\ref{powerabsorption}) and (\ref{matrixpowerabsorption}) 
\begin{equation}\label{Meijer-Inverse Wishart CFT}
\begin{split}
    \frac{1}{\Gamma_p({\bf \boldsymbol{\beta}} + \frac{3}{2}\boldsymbol{1})} \int &dY |Y|^{-1/2} \exp(-{\bf r}^\top A^{1/2} Y^{-1} A^{1/2} {\bf r} ) \\ \times &\; \overline{G}^{0, N}_{N, 0} \left( 
    \begin{array}{c}
    -{\bf \boldsymbol{\beta}} - \boldsymbol{1} \\
    -
    \end{array} \middle \vert Y
    \right) \\ =\frac{1}{\Gamma({\bf \boldsymbol{\beta}} + \frac{3}{2}\boldsymbol{1})}
        \int &dy y^{-1/2} \exp(-{\bf r}^\top A {\bf r}/y)  \\ \times &\; G^{0, N}_{N, 0} \left( 
    \begin{array}{c}
   -{\bf \boldsymbol{\beta}} - \boldsymbol{1}\\
    -
    \end{array} \middle \vert y 
    \right) 
\end{split}
\end{equation}
In order to obtain Eq.~(\ref{CFTInverse2}) we need to make $A = (\sqrt{2\omega}\Sigma_0)^{-1}$ and multiply both sides by $1/\Gamma_p (\boldsymbol{\beta} + \boldsymbol{1})$ and the definition in Eq.~(\ref{multivariategamma}).

\end{document}